\documentclass[twocolumn,superscriptaddress,letter]{revtex4}%
\usepackage{amssymb}
\usepackage{color}
\usepackage{graphicx}
\usepackage{dcolumn}
\usepackage{bm}
\usepackage[header,title,page,titletoc]{appendix}
\usepackage{amsmath}
\usepackage{amsfonts}%
\providecommand{\U}[1]{\protect \rule{.1in}{.1in}}

\begin{document}
\title{Majorana Edge States for $\mathrm{Z}_{2}$ Topological Orders of the
Wen-plaquette Model and the Toric-code Model}
\author{Jing Yu}
\affiliation{Department of Physics, Beijing Normal University, Beijing, 100875 P. R. China }
\affiliation{Department of Physics, Liaoning Shihua University, Fushun, 113001 P. R. China}
\author{Xing-Hai Zhang}
\affiliation{Department of Physics, Beijing Normal University, Beijing, 100875 P. R. China }
\author{Su-Peng Kou}
\thanks{Corresponding author}
\email{spkou@bnu.edu.cn}
\affiliation{Department of Physics, Beijing Normal University, Beijing, 100875 P. R. China }

\begin{abstract}
In this paper\emph{\ }we study the symmetry protected Majorana edge states for
the $\mathrm{Z}_{2}$ topological order of the Wen-plaquette model and the
toric-code model and calculate the dispersion of the Majorana edge states. For
the system with translational symmetry, the Majorana edge states are gapless
and have the nodal points at $k=0$ and $k=\pi$. For the edge states of the
toric-code model without translational symmetry, the edge modes become gapped.

\end{abstract}
\maketitle

\section{Introduction}

Topological properties of topologically ordered states can be partially
characterized by their gapless edge states\cite{W9038,W9125,int,wen}. For
example, the fractional quantum hall(FQH) states possess robust gapless edge
modes as protected by the energy gap of the bulk, which can be derived from
the Laughlin wave function\cite{la}. On the other hand, people can obtain the
bulk properties from the information of the edge states by bulk-boundary
correspondence. For example, edge excitation of FQH states is chiral Luttinger
liquid\cite{W9038,W9125} while that of non-Abelian FQH states\cite{mr} is more
exotic and related to (1+1)-dimensional conformal field
theories\cite{wen5,int}.

Recently, several exactly solvable spin models are found with non-Abelian
topological ordered state or $\mathrm{Z}_{2}$ topological ordered state as the
ground states, such as the toric-code model\cite{k1}, the Wen-plaquette
model\cite{wen,wen1} on a square lattice and the Kitaev model on a honeycomb
lattice\cite{k2}. In non-Abelian topological order of the Kitaev model on a
honeycomb lattice the elementary excitation becomes non-Abelian anyon with
nontrivial statistics. And the edge state is 1D gapless chiral Majorana modes.
However, for a simpler example of a topological order - $\mathrm{Z}_{2}$
topological order, gapless edge states have not been found while a gapped edge
state may exist\cite{bra,kong,bsw,chen} instead. In the toric-code model, it
is pointed out that there are two distinct types of boundaries (the smooth one
and the rough one) with gapped edge states\cite{bra,kong}. We call them all
zigzag boundary in this paper.

In this paper, instead of considering the zigzag boundaries that have gapped
edge states, we study the Wen-plaquette model and the toric-code model with a
new type of smooth boundary (which differs from the the smooth edge in
Ref.\cite{bra,kong}) and use Majorana formulation to derive the effective
theory of the gapless edge states.\ See the illustrations below. The gapless
edge states are protected by translational symmetry along the boundary. When
the translational symmetry is broken, the edge modes will have energy gap.

The paper is organized as follows. In Sec.II, we study the edge states of the
Wen-plaquette model. In this section, we give the definition\ of the edge
states by string operators and use the Majorana representation to describe
Majorana edge states. In Sec.III we study the edge states of the toric-code
model by similar approach. Finally, the conclusions are given in Sec.V.

\section{The edge states of the Wen-plaquette model}

\subsection{The Wen-plaquette model}

The Wen-plaquette model is defined on square lattice with the Hamiltonian
\begin{equation}
\hat{H}=-g\sum_{i}\hat{F}_{i},
\end{equation}
with
\begin{equation}
\hat{F}_{i}=\hat{\sigma}_{i}^{x}\hat{\sigma}_{i+\hat{e}_{x}}^{y}\hat{\sigma
}_{i+\hat{e}_{x}+\hat{e}_{y}}^{x}\hat{\sigma}_{i+\hat{e}_{y}}^{y}%
\end{equation}
and $g>0.$ $\hat{\sigma}_{i}^{x},$ $\hat{\sigma}_{i}^{y}$ are Pauli matrices
on site $i.$

The ground states of the Wen-plaquette model are known as $\mathrm{Z}_{2}$
topological state\cite{wen,wen1}. The ground state is denoted by ${F_{i}%
\equiv+1}$ at each plaquette. For this model, the elementary excitations are
$\mathrm{Z}_{2}$ vortex ($m$ type excitation denoted by ${F_{i}=-1}$ at even
sub-plaquette) and $\mathrm{Z}_{2}$ charge ($e$ type excitation denoted by
${F_{i}=-1}$ at odd sub-plaquette). In addition, there is a mutual-semion
statistics between $\mathrm{Z}_{2}$ vortex and $\mathrm{Z}_{2}$ charge. A
$\mathrm{Z}_{2}$ vortex and a $\mathrm{Z}_{2}$ charge annihilate with each
other into a fermionic $\mathrm{Z}_{2}$ link-excitation which is a pair of
$\mathrm{Z}_{2}$ vortex and $\mathrm{Z}_{2}$ charge. The fermions have flat
band - the energy spectrum is $E(\mathbf{k})=4g$, which implies that they
cannot move at all.

\begin{figure}[ptb]
\includegraphics[width=0.5\textwidth]{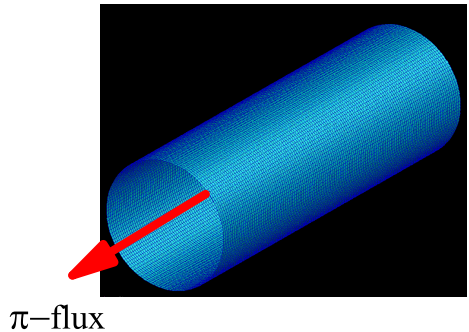}\caption{The illustration of a
system on a cylinder. The topological degeneracy arises from the presence or
the absence of $\pi$ flux of fermions through the hole.}%
\label{Fig.2}%
\end{figure}

The ground states of $\mathrm{Z}_{2}$ topological state have topological
degeneracy. Under the periodic boundary condition (on a torus), the ground
states of the Wen-plaquette model have four-fold degeneracy on even-by-even
($e\ast e$) lattice, two-fold degeneracy on even-by-odd, odd-by-even and
odd-by-odd lattices. For a system on a cylinder, the ground states have
two-fold degeneracy. Physically, the topological degeneracy arises from the
presence or the absence of $\pi$ flux of fermions through the hole, as
illustrated in Fig. \ref{Fig.2}.

\subsection{String representation of the edge states for the Wen-plaquette
model}

In Ref.\cite{kou2}, it is pointed out that for the Wen-plaquette model with
smooth open boundary condition, there exist gapless edge states. Fig.2 shows
the Wen-plaquette model with a smooth open boundary and Fig.3 shows the
Wen-plaquette model with a zigzag open boundary. For a finite $L_{x}\times
L_{y}$ lattice with a periodic boundary condition only along $y$-direction,
there are two edges along $y$-direction. This model can be realized by setting
$g=0$ for one column of plaquettes and it remains exactly solvable. The ground
states have $\sim2^{L_{y}}$-fold degeneracy and can be viewed as gapless edge
excitations on both boundaries described by Majorana fermion. These gapless
edge states can be mapped to a Majorana fermion system with flat band exactly.
Above argument comes from the exactly solvable model. But how the edge states
change when there exist external fields? From the effective $\mathrm{Z}_{2}$E
type mutual \textrm{U(1)}$\times$\textrm{U(1)} Chern-Simons(CS) theories of
$\mathrm{Z}_{2}$ topological order of the Wen-plaquette model, we have shown
that there are right-moving and left-moving gapless edge excitations described
by Majorana fermions, provided that the edge is in the $x$- or $y$%
-direction\cite{kou2}. The presence of the translational symmetry in the $x$-
or $y$-direction is crucial for the existence of the gapless edge excitations
for the $\mathrm{Z}2$E type mutual \textrm{U(1)}$\times$\textrm{U(1) }CS
theory and the lattice model. In addition, from the classification of the
$\mathrm{Z}_{2}$ topological order on a square lattice, we found that the
emergent Majorana edge states of the Wen-plaquette model will always gapless
at $k_{x}=0,$ $\pi$ (or $k_{y}=0,$ $\pi$) in momentum space\cite{kou3,kou4}.

\begin{figure}[ptb]
\includegraphics[width=0.5\textwidth]{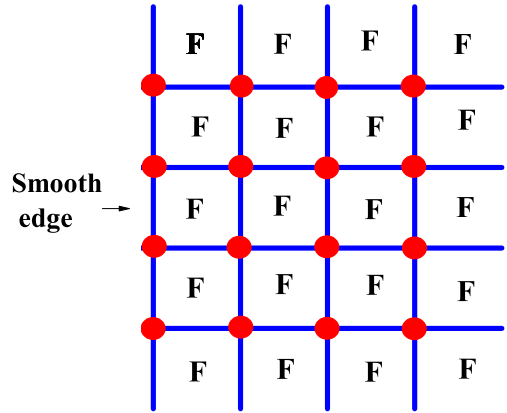}\caption{The illustration of a
smooth boundary of the Wen-plaquette model. The boundary has translational
symmetry.}%
\label{Fig.3}%
\end{figure}

\begin{figure}[ptb]
\includegraphics[width=0.5\textwidth]{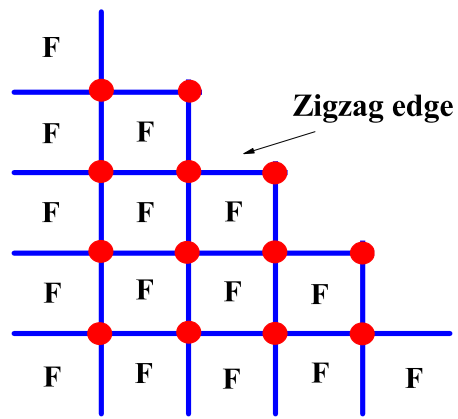}\caption{The illustration of a
zigzag boundary of the Wen-plaquette model. The boundary has no translational
symmetry.}%
\label{Fig.4}%
\end{figure}

To describe the edge states of the ground states (the planar codes), we define
the fermion string operators. In the bulk, the fermion string operators are
$\hat{W}_{f}(C)=\prod \limits_{m}\hat{\sigma}_{i_{m}}^{l_{m}}$\cite{wen2},
where $C$ is a string connecting the middle of the nearby sites, and $i_{m}$
are sites on the string. $l_{m}=z$ if the string does not turn at site $i_{m}%
$. $l_{m}=x$ or $y$ if the string makes a turn at site $i_{m}$. $l_{m}=y$ if
the turn forms a upper-right or lower-left corner. $l_{m}=x$ if the turn forms
a lower-right or upper-left corner. Taking the open boundary condition into
account, the loop $C$ can be different. For system with open boundary
condition, $C$ is a string from one boundary to another. It is obvious that
there are two kinds of strings, one has two ends at the same boundary (we call
it $C_{s}$), the other terminates at different boundaries (we call it $C_{l}%
$). See Fig.4.

Now we can create the edge states $\left \vert \mathrm{edge}\right \rangle $ by
performing a fermion string operation $\hat{W}_{f}(C_{s/l})$ connecting the
boundaries on the ground state $\left \vert 0\right \rangle $ as
\[
\left \vert \mathrm{edge}\right \rangle =\hat{W}_{f}(C_{s/l})\left \vert
0\right \rangle .
\]
Both the edge states $\left \vert \mathrm{edge}\right \rangle $ and the ground
states $\left \vert 0\right \rangle $ can be denoted by ${F_{i}\equiv+1}$ at
each plaquette and have the same ground state energy as
\[
E_{0}=\hat{H}_{\mathrm{w}}\left \vert \mathrm{edge}\right \rangle =H_{\mathrm{w}%
}\left \vert 0\right \rangle .
\]
It is obvious that for the Wen-plaquette model, both types of fermion strings
connecting the boundaries are condensed. These edge states have exact zero
energy or flat band.

\begin{figure}[ptb]
\includegraphics[width=0.5\textwidth]{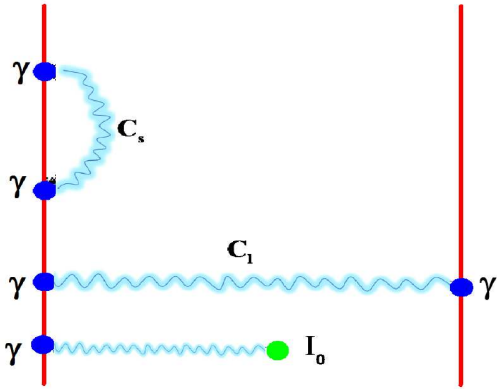}\caption{The illustration of
three fermion strings: the closed string $C_{s}$ corresponds to the string
with two ends at the same boundary; another closed string operator $C_{l}%
$\ corresponds to the string with two ends at the different boundaries; the
third corresponds to the open fermion string with one end (the blue spot) on
the boundary and the other (the green spot) in the bulk (site $I_{0}$ in the
bulk). There exists Majorana fermion mode on each end of the open string.}%
\label{Fig.5}%
\end{figure}

Let's explain why the edge states are really Majorana fermions. People know
that for an open fermion string, there exists a Majorana fermion modes at each
end. In Fig.4, we have shown an open fermion string, of which one end (the
blue spot) is on the boundary, the other (the green spot) is in the bulk (site
$I_{0}$ in the bulk). The end of fermion string in the bulk corresponds to a
Majorana fermion mode. Since the fermion parity of system is conserved, there
must exist another Majorana fermion mode at the boundary of the system which
is another end of the fermion string (site $i$ on the boundary). Thus each end
of the fermion string connecting the boundaries corresponds to a Majorana
fermion mode. On the other hand, each fermion string connecting the boundaries
corresponds to a two-level state denoted by $\hat{W}_{f}(C_{s/l})\left \vert
0\right \rangle =\pm \left \vert 0\right \rangle .$ Then we can say that each end
of the fermion string connecting the boundaries is really a Majorana fermion
mode. There are two types of fermion string operators, $\hat{W}_{f}(C_{s})$
denotes the string with two ends at the same boundary, $\hat{W}_{f}(C_{l})$
denotes the string with two ends at the different boundaries.

When we add external field terms
\begin{equation}
\hat{H}_{I}=h^{x}\sum \limits_{i}\hat{\sigma}_{i}^{x}+h^{y}\sum \limits_{i}%
\hat{\sigma}_{i}^{y}+h^{z}\sum \limits_{i}\hat{\sigma}_{i}^{z},
\end{equation}
the degeneracy of the edge states will be removed and we will get the
dispersive edge states. In the following parts we will calculate the
dispersion of these edge states and point out that for the Wen-plaquette model
with smooth boundary the gapless edge states are protected by the
translational symmetry.

In addition, for this case, the edge states on a finite $L_{x}\times L_{y}$
lattice with a periodic boundary condition only along $y$-direction always
have two-fold degeneracy. The two-fold degeneracy is characterized by the
closed string operator along $x$-direction $\hat{W}_{f}(C_{l})=\hat{\sigma
}_{i+\hat{e}_{x}}^{z}\hat{\sigma}_{i+2\hat{e}_{x}}^{z}\cdots \hat{\sigma
}_{i+L_{x}\hat{e}_{x}}^{z}$. For the case of $\hat{W}_{f}(C_{l})\left \vert
0\right \rangle =\left \vert 0\right \rangle ,$ there is no $\pi$-flux inside the
hole and the Majorana fermions on the edge have periodic boundary condition;
for the case of $\hat{W}_{f}(C_{l})\left \vert 0\right \rangle =-\left \vert
0\right \rangle ,$ there is a $\pi$-flux inside the hole and the Majorana
fermions on the edge have anti-periodic boundary condition.

\subsection{Majorana edge states for the Wen-plaquette model}

In the following parts we will study the (symmetry protected) edge states by
Majorana representation. Now we may use a Majorana mode $\gamma_{i}$ to denote
an end of the fermion string at boundary. For the boundary shown in Fig.2, the
corresponding effective Hamiltonian of a single edge mode is given by
\begin{equation}
\hat{H}_{\mathrm{edge}}=i\sum_{ij}J_{ij}\gamma_{i}\gamma_{j}%
\end{equation}
where $\gamma_{i}$ is the Majorana operators at edge position $i$, and obeys
algebra relation $\{ \gamma_{i},\gamma_{j}\}=\delta_{ij}$,\quad$(\gamma
_{i})^{\dag}=\gamma_{i}$.

\subsubsection{Quantum tunneling effect of Majorana modes}

Firstly, we select the Majorana edge states characterized by fermion string
operator $\hat{W}_{f}(C_{s})=\prod \limits_{m}\hat{\sigma}_{i_{m}}^{l_{m}}$. A
fermion string operator $\hat{W}_{f}(C_{s})$ that connects the two points on
boundary can be considered as quantum tunneling processes of virtual
quasi-particles moving along the path. The quantum tunneling process of
fermions is defined as : at first a single (bulk) fermion and an edge fermion
are created together. Then this bulk fermion propagates and disappears at the
boundary site $j$. And a string of $\hat{\sigma}_{i}^{m}$ is left on the
tunneling path behind the virtual fermion, that is just a string operator
$\hat{W}_{f}(C_{s})$. For simplicity, we use $\left \vert 0\right \rangle $ and
$\left \vert 1\right \rangle $ to describe the two degenerate eigenstates of the
string operator $\hat{W}_{f}(C_{s})$, as
\[
\hat{W}_{f}(C_{s})\left \vert 0\right \rangle =\left \vert 0\right \rangle ,\text{
}\hat{W}_{f}(C_{s})\left \vert 1\right \rangle =-\left \vert 1\right \rangle .
\]
With the perturbation term $\hat{H}_{I}$, the quantum tunneling processes
occur - the fermion propagates from one end of the string to the other. Thus
we can use the quantum tunneling theory in Ref.\cite{kou1,yu} to obtain the
hopping parameters for the Majorana modes, $J_{ij}$. The nearest neighbor
hopping parameter $J_{1}$ of the Majorana edge modes corresponds to the
shortest fermion string with two ends at the same boundary, of which the
fermion string operator is $\hat{W}_{f}(C_{s})=\hat{\sigma}_{i}^{z}\hat
{\sigma}_{i+\hat{e}_{x}}^{x}\hat{\sigma}_{i+\hat{e}_{x}+\hat{e}_{y}}^{y}%
\hat{\sigma}_{i+\hat{e}_{y}}^{z}$. See the illustration in Fig.5. And the next
nearest neighbor hopping parameter $J_{2}$ corresponds to a fermion string
with two ends at the same boundary, of which the fermion string operator is
$\hat{W}_{f}(C_{s})=\hat{\sigma}_{i}^{z}\hat{\sigma}_{i+\hat{e}_{x}}^{x}%
\hat{\sigma}_{i+\hat{e}_{x}+\hat{e}_{y}}^{z}\hat{\sigma}_{i+\hat{e}_{x}%
+2\hat{e}_{y}}^{y}\hat{\sigma}_{i+2\hat{e}_{y}}^{z}.$

\begin{figure}[ptb]
\includegraphics[width=0.5\textwidth]{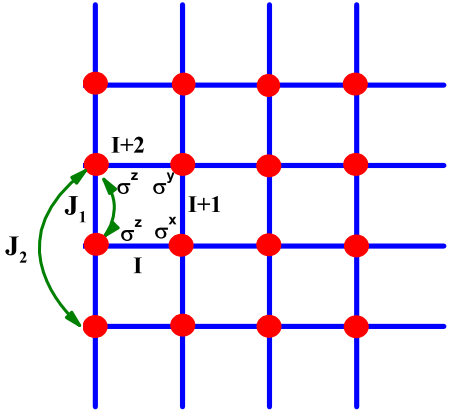}\caption{The illustration of
the relation between the hopping parameter $J_{1}$ in the effective model of
the edge states and the quantum tunneling process. There is translational
symmetry for the edge states along the boundary. }%
\label{Fig.6}%
\end{figure}

Let's calculate $J_{1}$ by the higher order perturbation approach. From the
higher order perturbation approach in Ref.\cite{kou1,yu}, we obtain the energy
shifts of two eigenstates $\left \vert 0\right \rangle $ and $\left \vert
1\right \rangle $. We take the energy shifts of the quantum state $\left \vert
1\right \rangle $ as an example which is
\begin{equation}
\delta E_{1}=\left \langle 1\right \vert \hat{H_{I}}(\frac{1}{E_{0}-\hat{H}_{0}%
}\hat{H_{I}})^{L_{0}-1}\left \vert 1\right \rangle .
\end{equation}
To calculate $(\frac{1}{E_{0}-\hat{H}_{0}}\hat{H_{I}})\left \vert
1\right \rangle ,$ we can choose site $i$ as the starting point of the
generation of an edge fermion mode and get
\begin{align}
(\frac{1}{E_{0}-\hat{H}_{0}}\hat{H_{I}})\left \vert 1\right \rangle  &
\rightarrow(\frac{h^{z}}{E_{0}-\hat{H}_{0}}\hat{\sigma}_{i}^{z})\left \vert
1\right \rangle \\
&  =(\frac{h^{z}}{E_{0}-\hat{H}_{0}})\left \vert \Psi_{i}\right \rangle
\nonumber
\end{align}
where $\left \vert \Psi_{i}\right \rangle $ is the excited state of an edge
fermion mode and a bulk fermion at link $I$ (See Fig.5). From $\hat{H}%
_{0}\left \vert \Psi_{i}\right \rangle =(E_{0}+4g)\left \vert \Psi_{i}%
\right \rangle ,$ we have%
\[
(\frac{1}{E_{0}-\hat{H}_{0}}\hat{H_{I}})\left \vert 1\right \rangle
=(\frac{h^{z}}{-4g})\left \vert \Psi_{i}\right \rangle .
\]
Then the bulk fermion move from $I$-link to $I+1$-link by an $\hat{\sigma
}_{i+\hat{e}_{x}}^{x}$ operation, then turn around to $I+2$-link by an
$\hat{\sigma}_{i+\hat{e}_{x}+\hat{e}_{y}}^{y}$ operation. Finally, the bulk
fermion disappears by performing the operation at $I+2$-link by $\hat{\sigma
}_{i+\hat{e}_{y}}^{z}$ operation at site $i+1$.

Then we can get the energy shift of the state $\left \langle 1\right \vert $ as%
\begin{equation}
\delta E_{1}=\sum \limits_{j=0}^{\infty}\left \langle 1\right \vert \hat{H}%
_{I}(\frac{1}{E_{0}-\hat{H}_{0}}\hat{H}_{I})^{j}\left \vert 1\right \rangle
=\frac{h^{x}\left(  h^{z}\right)  ^{2}h^{y}}{(-4g)^{3}}.
\end{equation}
Using the same approach we can get the energy shift of the state $\left \langle
0\right \vert $ as%
\begin{equation}
\delta E_{0}=\sum \limits_{j=0}^{\infty}\left \langle 0\right \vert \hat{H}%
_{I}(\frac{1}{E_{0}-\hat{H}_{0}}\hat{H}_{I})^{j}\left \vert 0\right \rangle
=-\frac{h^{x}\left(  h^{z}\right)  ^{2}h^{y}}{(-4g)^{3}}.
\end{equation}
Finally an energy difference $\varepsilon$ of the two quantum states is
obtained as
\begin{equation}
\varepsilon=\delta E_{1}-\delta E_{0}=\frac{2h^{x}\left(  h^{z}\right)
^{2}h^{y}}{(-4g)^{3}}%
\end{equation}
which is the strength of the nearest neighbor hopping parameter $J_{1}$ of the
Majorana edge modes,
\[
\varepsilon=J_{1}=\frac{2h^{x}\left(  h^{z}\right)  ^{2}h^{y}}{(-4g)^{3}}.
\]
Similarly we can derive the next nearest neighbor hopping parameter $J_{2}$ of
the Majorana edge modes by using the same approach as
\begin{equation}
J_{2}=\frac{2h^{x}\left(  h^{z}\right)  ^{3}h^{y}}{(-4g)^{4}}.
\end{equation}

\subsubsection{Symmetry protected Majorana edge states}

In this part, we use the Majorana representation to derive the dispersion of
the Majorana edge states. By the following representation,
\[
\gamma_{i}=\frac{1}{\sqrt{2}}\left(  c_{i}+c_{i}^{\dag}\right)  \, ,
\]
we have
\begin{align}
\hat{H}_{\mathrm{edge}}  &  =i\sum_{\left \langle ij\right \rangle }J_{1}
\gamma_{i}\gamma_{j}+i\sum_{\left \langle \left \langle ij\right \rangle
\right \rangle }J_{2}\gamma_{i}\gamma_{j}+...\\
&  =\frac{iJ_{1}}{2}\sum_{i}(c_{i}+c_{i}^{\dag})(c_{i+1}+c_{i+1}^{\dag
})\nonumber \\
&  +\frac{iJ_{2}}{2}\sum_{i}(c_{i}+c_{i}^{\dag})(c_{i+2}+c_{i+2}^{\dag
})+...\nonumber \\
&  =\frac{iJ_{1}}{2}\sum_{i}[c_{i}^{\dag}c_{i+1}-c_{i+1}^{\dag}c_{i}%
+c_{i}^{\dag}c_{i+1}^{\dagger}+c_{i}c_{i+1}]\nonumber \\
&  +\frac{iJ_{2}}{2}\sum_{i}[c_{i}^{\dag}c_{i+2}-c_{i+2}^{\dag}c_{i}%
+c_{i}^{\dag}c_{i+2}^{\dagger}+c_{i}c_{i+2}]\nonumber \\
&  +...\nonumber
\end{align}

In the momentum space we derive the energy spectrum of the edge states as
\begin{align}
E_{+}  &  \simeq2(J_{1}\sin k+J_{2}\sin2k),\\
E_{-}  &  =0.\nonumber
\end{align}
Here the quantum states with zero energy are un-physical. Thus the excited
energy of the edge states is given by%
\[
\Delta E=\left \vert E_{+}\right \vert =2\left \vert J_{1}\sin k+J_{2}%
\sin2k\right \vert \,.
\]
We found that the nodal points of $\mathrm{Z}_{2}$\textit{\ }topological order
are fixed at $k=0$ and $k=\pi$ on an edge which is protected by the
$\mathrm{Z}_{2}$ topological invariants and translational invariance. See Fig.6.

\begin{figure}[ptb]
\includegraphics[width=0.5\textwidth]{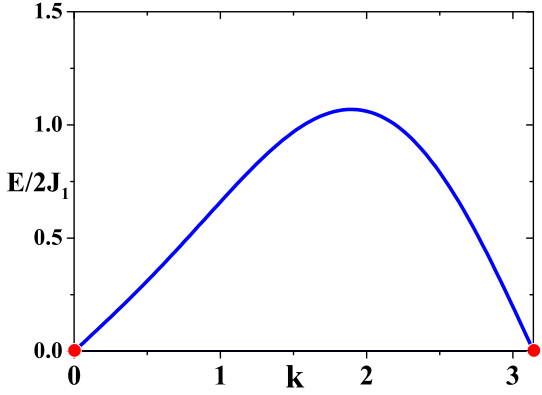}\caption{The excited energies
of a single edge mode of the Wen-plaquette model. The nodal points of
$\mathrm{Z}_{2}$\textit{\ }topological order are fixed at $k=0$ and $k=\pi$
(the red spots). We have set $J_{2}=-0.2J_{1}$.}%
\label{Fig.7}%
\end{figure}

\subsubsection{Interference between the Majorana fermions on two boundaries}

In this part we study the interference between the Majorana fermions on two
boundaries. Now the effective Hamiltonian of two coupling Majorana edge modes
is given by
\begin{align}
H_{\mathrm{edge}}  &  \simeq i\sum_{i}J_{t}\gamma_{A,i}\gamma_{B,i}\nonumber \\
&  +i\sum_{i}J_{1}\gamma_{A,i}\gamma_{A,i+1}+i\sum_{i}J_{1}\gamma_{B,i}%
\gamma_{B,i+1}\nonumber \\
&  +i\sum_{i}J_{2}\gamma_{A,i}\gamma_{A,i+1}+i\sum_{i}J_{2}\gamma_{B,i}%
\gamma_{B,i+1}%
\end{align}
where $A$ and $B$ denote the indices of the two boundaries. $J_{t}$ represents
the coupling strength between two boundaries.

To derive $J_{t}$, we need to consider another type of string operators as
$\hat{W}_{f}(C_{l})=\hat{\sigma}_{i+\hat{e}_{y}}^{z}\hat{\sigma}_{i+2\hat
{e}_{y}}^{z}\cdots \hat{\sigma}_{i+L_{y}\hat{e}_{y}}^{z}$. Such string operator
is described by the quantum tunneling process of a link-fermion moving from
one boundary to another along y-direction. The energy difference of the double
states is
\begin{equation}
\varepsilon=-8g(\frac{h^{z}}{-4g})^{L_{y}}\rightarrow J_{t}.
\end{equation}
Now $L_{y}$ is not very big.

In addition, for a finite system we need to consider the quantum tunneling
effect along $x$-direction which will remove the two-fold degeneracy of all
quantum states including the edge modes. Because such quantum tunneling effect
is characterized by the closed string operator $\hat{W}_{f}(C_{l})=\hat
{\sigma}_{i+\hat{e}_{x}}^{z}\hat{\sigma}_{i+2\hat{e}_{x}}^{z}\cdots \hat
{\sigma}_{i+L_{x}\hat{e}_{x}}^{z}$ along x-direction, the energy splitting is
about $-8g(\frac{h^{z}}{-4g})^{L_{x}}$. Here we consider the case with big
$L_{x}$. Thus we can ignore this quantum tunneling effect.

For this case of $\hat{W}_{f}(C_{l})\left \vert 0\right \rangle =\left \vert
0\right \rangle ,$ there is no $\pi$-flux inside the hole and the Majorana
edges have periodic boundary condition. From $\gamma_{A,i}=\left(
c_{A,i}+c_{A,i}^{\dag}\right)  /\sqrt{2}$ and $\gamma_{B,i}=\left(
c_{B,i}+c_{B,i}^{\dag}\right)  /\sqrt{2}$, we have%
\begin{align}
\hat{H}_{\mathrm{edge}}  &  \simeq i\sum_{i}J_{t}\gamma_{A,i}\gamma
_{B,i}\nonumber \\
&  +i\sum_{i}J_{1}\gamma_{A,i}\gamma_{A,i+1}+i\sum_{i}J_{1}\gamma_{B,i}%
\gamma_{B,i+1}\nonumber \\
&  +i\sum_{i}J_{2}\gamma_{A,i}\gamma_{A,i+1}+i\sum_{i}J_{2}\gamma_{B,i}%
\gamma_{B,i+1}\nonumber \\
&  =\frac{iJ_{t}}{2}\sum_{i}\left(  c_{A,i}+c_{A,i}^{\dag}\right)  \left(
c_{B,i}+c_{B,i}^{\dag}\right) \nonumber \\
&  +\frac{iJ_{1}}{2}\sum_{i}\left(  c_{A,i}+c_{A,i}^{\dag}\right)  \left(
c_{A,i+1}+c_{A,i+1}^{\dag}\right) \nonumber \\
&  +\frac{iJ_{1}}{2}\sum_{i}\left(  c_{B,i}+c_{B,i}^{\dag}\right)  \left(
c_{B,i+1}+c_{B,i+1}^{\dag}\right) \nonumber \\
&  +\frac{iJ_{2}}{2}\sum_{i}\left(  c_{A,i}+c_{A,i}^{\dag}\right)  \left(
c_{A,i+2}+c_{A,i+2}^{\dag}\right) \nonumber \\
&  +\frac{iJ_{2}}{2}\sum_{i}\left(  c_{B,i}+c_{B,i}^{\dag}\right)  \left(
c_{B,i+2}+c_{B,i+2}^{\dag}\right)  .
\end{align}
In the momentum space we have
\begin{equation}
H_{\mathrm{edge}}=\sum_{k>0}(c_{A,k},c_{A,-k}^{\dagger},c_{B,k},c_{B,-k}%
^{\dagger})\epsilon(k)\left(
\begin{array}
[c]{c}%
c_{A,k}^{\dagger}\\
c_{A,-k}\\
c_{B,k}^{\dagger}\\
c_{B,-k}%
\end{array}
\right)
\end{equation}
where
\begin{equation}
\epsilon(k)=\left(
\begin{array}
[c]{cccc}%
\varepsilon_{k} & \varepsilon_{k} & iJ_{t}/2 & iJ_{t}/2\\
\varepsilon_{k} & \varepsilon_{k} & iJ_{t}/2 & iJ_{t}/2\\
-iJ_{t}/2 & -iJ_{t}/2 & \varepsilon_{k} & \varepsilon_{k}\\
-iJ_{t}/2 & -iJ_{t}/2 & \varepsilon_{k} & \varepsilon_{k}%
\end{array}
\right)
\end{equation}
and $\varepsilon_{k}=J_{1}\sin k+J_{2}\sin2k$. Now we have the eigenvalues as
\begin{equation}
E=0,0,2\left(  J_{1}\sin k+J_{2}\sin{2k}\right)  \pm J_{t}\,.
\end{equation}
In particular, for periodic boundary condition. The momentum is
\[
k=\frac{2\pi n}{M},\text{ }n=1,2...,M
\]
where $M$ is an integer number.

\begin{figure}[ptb]
\includegraphics[width=0.5\textwidth]{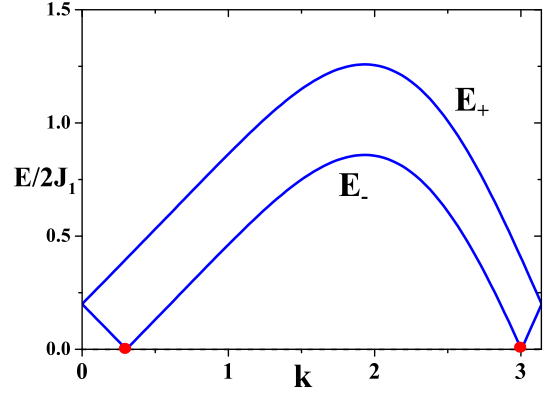}\caption{The excited energies
of two edge modes of the Wen-plaquette model with interference effect. The
nodal points of $\mathrm{Z}_{2}$\textit{\ }topological order shift away from
$k=0$ and $k=\pi$ (the red spots). We have set $J_{2}=-0.2J_{1},$
$J_{t}=0.2J_{1}.$}%
\label{Fig.8}%
\end{figure}

\begin{figure}[ptb]
\includegraphics[width=0.5\textwidth]{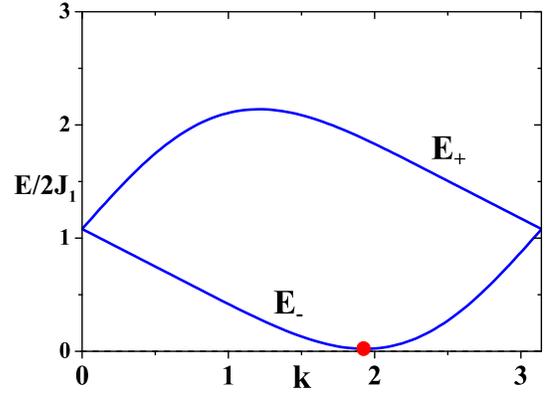}\caption{The excited energies
of two edge modes of the Wen-plaquette model with interference effect. There
exists a gapless point with quadratic dispersion (the red spot). We have set
$J_{2}=-0.2J_{1},$ $J_{t}=1.08J_{1}.$}%
\label{Fig.9}%
\end{figure}

\begin{figure}[ptb]
\includegraphics[width=0.5\textwidth]{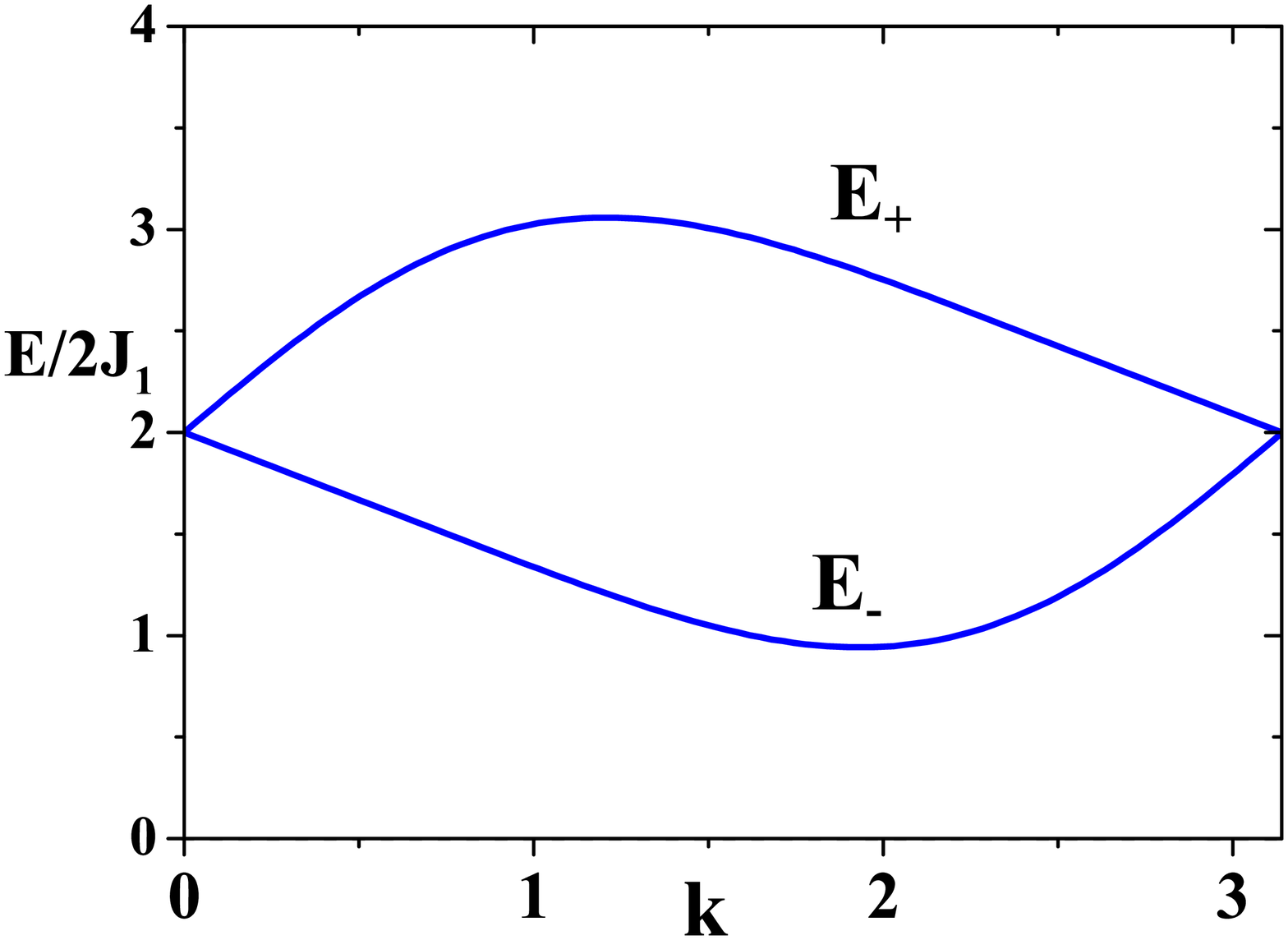}\caption{The excited energies
of two edge modes of the Wen-plaquette model with interference effect. The
edge states are full gapped. We have set $J_{2}=-0.2J_{1},$ $J_{t}=2J_{1}.$}%
\label{Fig.10}%
\end{figure}

On the other hand, for the case of $W(C_{l})\left \vert 0\right \rangle
=-\left \vert 0\right \rangle ,$ there is a $\pi$-flux inside the hole and the
Majorana edges have anti-periodic boundary condition. We have similar energy
levels. However due to the anti-periodic boundary condition, the momentum is
\[
k=\frac{2\pi n-\pi}{M},\text{ }n=1,2...,M
\]
where $M$ is an integer number. The two cases are always degenerate for an
infinite system.

Because the quantum states with zero energy are un-physical, the excited
energies are
\begin{align}
E_{+}  &  =\left \vert 2\left(  J_{1}\sin k+J_{2}\sin{2k}\right)
+J_{t}\right \vert ,\nonumber \\
E_{-}  &  =\left \vert 2\left(  J_{1}\sin k+J_{2}\sin{2k}\right)
-J_{t}\right \vert .
\end{align}
We found that the nodal points of $\mathrm{Z}_{2}$\textit{\ }order are not
fixed at $k=0/\pi$ on an edge due to the interference effect between the
Majorana fermions on two boundaries. See Fig.7. For $J_{t}>1.08J_{1},$ the
edge states become full gapped by mixing the edge states on different
boundaries. See Fig.9.

In particular, for a special case $\hat{H}_{I}=h^{z}\sum \limits_{i}\hat
{\sigma}_{i}^{z}$, the bulk fermion can only move straightforwardly but cannot
turn a corner. Now we have $J_{1}=J_{2}=0$. Then the effective Hamiltonian of
two coupling Majorana edge modes becomes
\begin{equation}
\hat{H}_{\mathrm{edge}}=iJ_{t}\sum_{i}\gamma_{A,i}\gamma_{B,j}.
\end{equation}
By the following representation of Majorana modes,
\[
\gamma_{A,i}=\left(  c_{i}+c_{i}^{\dag}\right)  /\sqrt{2},\qquad \gamma
_{B,i}=\left(  c_{i}-c_{i}^{\dag}\right)  /(\sqrt{2}i),
\]
we have
\begin{align*}
\hat{H}_{\mathrm{edge}}  &  =iJ_{t}\sum_{i}(c_{i}+c_{i}^{\dag})(-i)(c_{i}%
-c_{i}^{\dag})\\
&  =J_{t}\sum_{i}c_{i}^{\dag}c_{i}.
\end{align*}
This is a complex fermion system with two energy levels.

\section{The edge states of the toric-code model}

\begin{figure}[ptb]
\includegraphics[width=0.5\textwidth]{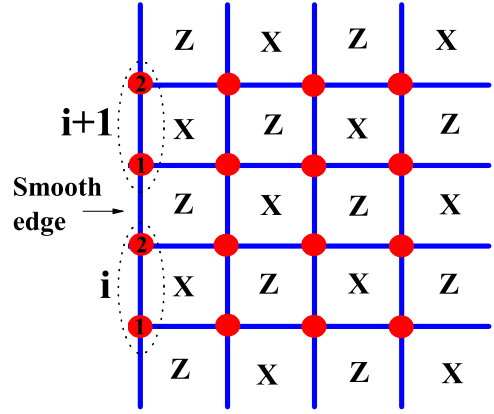}\caption{The illustration of a
smooth boundary of the toric-code model. The boundary has translational
symmetry for the case of $A=B$. When there is no translational symmetry for
the case of $A\neq B$, the unit-cell along the edge has two sites. Here we
denote the unit-cells by $i$ and $i+1$.}%
\label{Fig.12}%
\end{figure}

The toric-code model is described by the Hamiltonian \cite{k1}
\begin{align}
\hat{H}_{\mathrm{tc}} &  =-A\sum \limits_{i\in \mathrm{even}}{\hat{Z}_{i}}%
-B\sum \limits_{i\in \mathrm{odd}}{\hat{X}_{i}+}\label{tc}\\
&  h^{x}\sum \limits_{i}\hat{\sigma}_{i}^{x}+h^{y}\sum \limits_{i}\hat{\sigma
}_{i}^{y}+h^{z}\sum \limits_{i}\hat{\sigma}_{i}^{z}.\nonumber
\end{align}
where
\[
{\hat{Z}_{i}=}\hat{\sigma}_{i}^{z}\hat{\sigma}_{i+\hat{e}_{x}}^{z}\hat{\sigma
}_{i+\hat{e}_{x}+\hat{e}_{y}}^{z}\hat{\sigma}_{i+\hat{e}_{y}}^{z},\text{
}{\hat{X}_{i}=}\hat{\sigma}_{i}^{x}\hat{\sigma}_{i+\hat{e}_{x}}^{x}\hat
{\sigma}_{i+\hat{e}_{x}+\hat{e}_{y}}^{x}\hat{\sigma}_{i+\hat{e}_{y}}^{x}.
\]
$\hat{\sigma}_{i}^{x,y,z}$ are Pauli matrices on sites, $i.$ In this paper we
only consider the case of $A>0$, $B>0.$ The ground state of the toric-code
model is also $\mathrm{Z}_{2}$ topological state which is similar to that of
the Wen-plaquette model. For the toric-code model, if we choose $A>0$, $B>0,$
the ground state of it is denoted by ${Z_{i}\equiv+1}$ and ${X_{i}\equiv+1}$
at each plaquette.

\subsection{String representation of the edge states for the toric-code model}

To describe the edge states of the ground states (the planar codes), we also
define the fermion string operators. In the bulk, the fermion string operators
are also $\hat{W}_{f}(C)=\prod \limits_{m}\hat{\sigma}_{i_{m}}^{l_{m}}$, where
$C$ is a string, and $i_{m}$ are sites on the string. $l_{m}=y$ if the string
does not turn at site $i_{m}$. $l_{m}=x$ or $z$ if the string makes a turn at
site $i_{m}$. $l_{m}=x$, if the string turns around the $X$-plaquette;
$l_{m}=z$, if the string turns around the $Z$-plaquette. For system with open
boundary condition, $C$ is a string from one boundary to another. It is
obvious that the two types of string operators correspond to the string with
two ends at the same boundary (we call it $C_{s}$) and that with two ends at
different boundaries (we call it $C_{l}$), respectively. See Fig.5. Similarly
we can create the edge states $\left \vert \mathrm{edge}\right \rangle $ by
doing a fermion string operation $\hat{W}_{f}(C_{s/l})$ connecting the
boundaries on the ground state $\left \vert 0\right \rangle $ as
\[
\left \vert \mathrm{edge}\right \rangle =\hat{W}_{f}(C_{s/l})\left \vert
0\right \rangle .
\]

\subsection{Majorana edge states for the toric-code model}

\begin{figure}[ptb]
\includegraphics[width=0.5\textwidth]{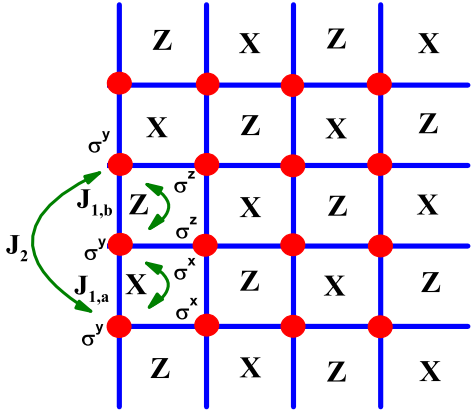}\caption{The illustration of
the relation between hopping parameters $J_{1,a},$ $J_{1,b}$ in the effective
model of the edge states for the toric-code model and quantum tunneling
processes.}%
\label{Fig.13}%
\end{figure}

\begin{figure}[ptb]
\includegraphics[width=0.5\textwidth]{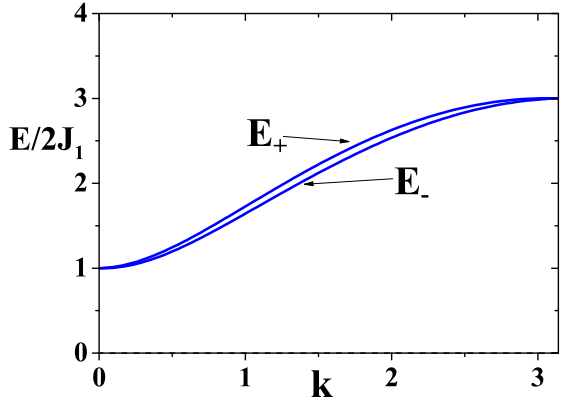}\caption{The excited energies
of a single edge mode of the toric-code model. The edge states are full gapped
without translational symmetry. We have set $J_{2}=-0.0125J_{1,a}$ and
$J_{1,b}=0.5J_{1,a}$. The spectrum splitting comes from a small next nearest
neighbor hopping.}%
\label{Fig.14}%
\end{figure}

Now we may use a Majorana mode $\gamma_{i}$ to denote an end of the fermion
string at boundary for the toric-code model. For the boundary shown in Fig.10,
the boundary has translational symmetry for the case of $A=B$. When there is
no translational symmetry for the case of $A\neq B$, the unit-cell along the
edge has two sites. Here we denote the unit-cells by $i$ and $i+1$. Now we
have two-sublattice in a unit-cell and denote them by $1$ and $2$. The
corresponding effective Hamiltonian of a single edge mode is given by
\begin{align}
\hat{H}_{\mathrm{edge}}  &  =iJ_{1,a}\sum_{i}\gamma_{1,i}\gamma_{2,i}%
+iJ_{1,b}\sum_{i}\gamma_{2,i}\gamma_{1,i+1}\nonumber \\
&  +iJ_{2}\sum_{i}\gamma_{1,i}\gamma_{1,i+1}+iJ_{2}\sum_{i}\gamma_{2,i}%
\gamma_{2,i+1}...
\end{align}
The nearest neighbor hopping parameters $J_{1,a}$ and $J_{1,b}$ of the
Majorana edge modes correspond to the fermion string operators $\hat{W}%
_{f}(C_{s})=\hat{\sigma}_{i}^{y}\hat{\sigma}_{i+\hat{e}_{x}}^{x}\hat{\sigma
}_{i+\hat{e}_{x}+\hat{e}_{y}}^{x}\hat{\sigma}_{i+\hat{e}_{y}}^{y}$ and
$\hat{W}_{f}(C_{s})=\hat{\sigma}_{i}^{y}\hat{\sigma}_{i+\hat{e}_{x}}^{z}%
\hat{\sigma}_{i+\hat{e}_{x}+\hat{e}_{y}}^{z}\hat{\sigma}_{i+\hat{e}_{y}}^{y}$,
respectively. See the illustration in Fig.11. And the next nearest neighbor
hopping parameter $J_{2}$ corresponds the fermion string operator $\hat{W}%
_{f}(C_{s})=\hat{\sigma}_{i}^{y}\hat{\sigma}_{i+\hat{e}_{x}}^{x}\hat{\sigma
}_{i+\hat{e}_{x}+\hat{e}_{y}}^{y}\hat{\sigma}_{i+\hat{e}_{x}+2\hat{e}_{y}}%
^{z}\hat{\sigma}_{i+2\hat{e}_{y}}^{y}.$ Similarly we can derive the nearest
neighbor hopping parameter $J_{1}$ and the next nearest neighbor hopping
parameter $J_{2}$ of the Majorana edge modes by using the same approach in
above section as
\begin{align}
J_{1,a}  &  =\frac{2h^{x}\left(  h^{y}\right)  ^{2}h^{x}}{(-2A-2B)^{3}},\text{
}J_{1,b}=\frac{2h^{z}\left(  h^{y}\right)  ^{2}h^{z}}{(-2A-2B)^{3}}\nonumber \\
J_{2}  &  =\frac{2h^{x}\left(  h^{y}\right)  ^{3}h^{z}}{(-2A-2B)^{4}}.
\end{align}
See Fig.11.

By defining $\gamma_{1,i}=(\psi_{1,i}+\psi_{1,i}^{\dag})/\sqrt{2}$,
$\gamma_{2,i}=(\psi_{2,i}-\psi_{2,i}^{\dag})/(\sqrt{2}i),$ we have
\begin{widetext}
\begin{equation*}
H_{edge}=\frac{1}{2}\left(
\begin{array}{cccc}
\psi _{A,k} & \psi _{A,-k}^{\dag } & \psi _{B,k} & \psi _{B,-k}^{\dag }%
\end{array}%
\right) \left(
\begin{array}{cccc}
2J_{2}\sin k & 2J_{2}\sin k & -J_{1,a}+J_{1,b}e^{ik} & J_{1,a}-J_{1,b}e^{ik}
\\
2J_{2}\sin k & 2J_{2}\sin k & -J_{1,a}+J_{1,b}e^{ik} & J_{1,a}-J_{1,b}e^{ik}
\\
-J_{1,a}+J_{1,b}e^{-ik} & -J_{1,a}+J_{1,b}e^{-ik} & 2J_{2}\sin k & -2J_{2}\sin k
\\
J_{1,a}-J_{1,b}e^{-ik} & J_{1,a}-J_{1,b}e^{-ik} & -2J_{2}\sin k & 2J_{2}\sin k%
\end{array}%
\right) \left(
\begin{array}{c}
\psi _{A,k}^{\dag } \\
\psi _{A,-k} \\
\psi _{B,k}^{\dagger } \\
\psi _{B,-k}%
\end{array}%
\right) .
\end{equation*}%
\end{widetext}The excited energies of the edge states are
\begin{align*}
E_{+}  &  =\left \vert 2J_{2}\sin k+\sqrt{J_{1,a}^{2}+J_{1,b}^{2}%
-2J_{1,a}J_{1,b}\cos k}\right \vert ,\\
E_{-}  &  =\left \vert 2J_{2}\sin k-\sqrt{J_{1,a}^{2}+J_{1,b}^{2}%
-2J_{1,a}J_{1,b}\cos k}\right \vert .
\end{align*}
If $J_{1,a}=J_{1,b}$ (or $A=B$), the system has translational symmetry and the
edge modes becomes gapless. In general, due to $J_{2}<<J_{1,a},J_{1,b},$ we
have the full gapped edge modes. In Fig.12, we set $J_{2}=0.0125J_{1,a}$ and
$J_{1,b}=0.5J_{1,a}$.

\section{Conclusion}

In this paper\emph{\ }we study the symmetry protected Majorana edge states for
the $\mathrm{Z}_{2}$ topological order of the Wen-plaquette model and those of
the toric-code model and calculate the dispersion of the Majorana edge states.
We found that for the Wen-plaquette model the nodal points are fixed at $k=0$
and $k=\pi$ on an edge which is protected by the $\mathrm{Z}_{2}$ topological
invariants and translational invariance. Due to the interference effect
between the Majorana fermions on two boundaries, the nodal points of the
Wen-plaquette model are not fixed at $k=0/\pi$. For strong interference case,
the edge states become full gapped. While, for the toric-code model with
$A\neq B$, there is no translational symmetry along the edges. Then the edge
modes becomes gapped. If we recover the translational symmetry by setting
$A=B,$ the edge states of the toric-code model have same properties to those
of the Wen-plaquette model.

\acknowledgments This work is supported by NFSC Grant No. 11174035, National
Basic Research Program of China (973 Program) under the grant No.
2011CB921803, 2012CB921704.

\end{document}